\documentclass[conference]{IEEEtran}

\usepackage{cite}
\usepackage{mdframed}
\usepackage{xcolor}

\ifCLASSINFOpdf
  \usepackage[pdftex]{graphicx}
\else
  \usepackage[dvips]{graphicx}
\fi
\usepackage{amsmath}
\usepackage{algorithmic}
\ifCLASSOPTIONcompsoc
 \usepackage[caption=false,font=normalsize,labelfont=sf,textfont=sf]{subfig}
\else
 \usepackage[caption=false,font=footnotesize]{subfig}
\fi
\usepackage{url}
\usepackage{booktabs}

\begin{document}
\title{Beyond End-to-End Success: Diagnosing Failures in Long-Horizon Security LLM Agents}

\author{
\IEEEauthorblockN{
    Wei Shao\IEEEauthorrefmark{1},
    Chongzhou Fang\IEEEauthorrefmark{2},
    Zuxiong Tan\IEEEauthorrefmark{1},
    Zequan Liang\IEEEauthorrefmark{1},
    Setareh Rafatirad\IEEEauthorrefmark{1},
    Avesta Sasan\IEEEauthorrefmark{1}, \\and
    Houman Homayoun\IEEEauthorrefmark{1}
}
\\
\IEEEauthorblockA{\IEEEauthorrefmark{1}University of California, Davis, USA\\
Email: \{wayshao, zxtan, zqliang, srafatirad, asasan, hhomayoun\}@ucdavis.edu}
\IEEEauthorblockA{\IEEEauthorrefmark{2}Rochester Institute of Technology, USA\\
Email: cxfeec@rit.edu}
}

\maketitle

\begin{abstract}
Long-horizon security LLM agents must carry information and decisions across many dependent interactions, where later actions often depend on services, state, or access discovered much earlier. This makes final task success difficult to interpret: an agent may fail before it ever reaches the point where the capability of interest can be exercised. We present a diagnostic methodology that instruments security tasks with checkpoints, separates failures before and after capability exposure, and uses controlled interventions to test suspected upstream bottlenecks. We evaluate the methodology across four task families involving delayed reuse of discovered information, reuse of observed state, recovery from failed strategies, and decision making after uncertain outcomes. On observed state reuse, checkpoint analysis shows that many Gemini 2.5 Flash failures occur before the model observes the state it is later expected to reuse. In a pre-specified 92-seed study, targeted protocol-disambiguation guidance increases state observation from 65.5\% under a matched non-guidance control message to 95.4\%. Repeating the same design with Gemini 3.7 Flash produces the opposite effect, while state observation no longer reliably predicts task completion. These results show that the dominant source of failure can shift across model generations, motivating evaluation that diagnoses where and why long-horizon security agents fail rather than relying only on aggregate task success.
\end{abstract}

% \fi
\IEEEpeerreviewmaketitle

\section{Introduction}
\label{sec:introduction}

Large language model (LLM) agents have demonstrated increasing capability in interacting with external tools and environments to execute complex, multi-step security tasks. Recent research has applied LLM agents to penetration testing, web exploitation, vulnerability discovery, and Capture-the-Flag (CTF) challenges~\cite{deng2024pentestgpt,fang2024hackwebsites,fang2024oneday,zhu2024zeroday,abramovich2025enigma,udeshi2025dcipher}. Concurrently, a growing body of work has introduced benchmarks that systematically evaluate these capabilities in controlled and progressively more realistic security environments~\cite{zhang2025cybench,gioacchini2025autopenbench,zhu2025cvebench,wang2025cybergym,mai2025shell,liu2026agentcyberrange}. As these agents continue to advance, understanding not only whether they succeed, but also where and why failures occur, has become increasingly important.

This analysis is especially critical for long-horizon tasks, where completing a goal requires many dependent interactions with the environment. A security agent may need to discover services, obtain information, use that information later, adapt when an action fails, and eventually reach the final objective. Reliability becomes increasingly important as these dependencies grow~\cite{kwa2025longtasks}, and interactive-agent benchmarks show that models continue to struggle over extended trajectories~\cite{liu2023agentbench,zhou2023webarena,yao2024taubench}. Existing cybersecurity benchmarks already start to consider factors beyond final task success. Some use subtasks or milestones to show intermediate progress, while others evaluate different stages of a security workflow or distinguish progressively stronger capabilities~\cite{yang2023intercode,shao2024nyuctf,zhang2025cybench,gioacchini2025autopenbench,yang2025pentesteval,lee2026exploitbench,wang2025cybergym,liu2026agentcyberrange}. Broader agent evaluations similarly measure partial progress or analyze failures throughout a trajectory~\cite{gioacchini2024agentquest,ma2024agentboard,gioacchini2025autopenbench,yang2025pentesteval,lee2026exploitbench,mazaheri2026agentatlas,wang2026horizon,jang2026odysseys}. However, intermediate progress alone does not tell us whether the agent actually exercised the capability a task was intended to test. For example, if a task is designed to test whether an agent can remember and later reuse a piece of state, a run that never discovers that state does not provide evidence of a memory or reuse failure. End-to-end failure therefore combines two different cases: failure to reach the point where the capability becomes relevant and failure after that point is reached.

In this work, we develop a diagnostic methodology for long-horizon security-agent evaluation. We instrument each task with checkpoints and identify the point at which the capability under study first becomes exercisable. We refer to reaching this point as \emph{exposure}. By reporting both whether the agent reaches exposure and what happens afterward, we distinguish upstream failures from failures of the intended capability itself. When checkpoint analysis suggests a specific upstream obstacle, we test that explanation using matched interventions:

\begin{itemize}
    \item \textbf{Rescue:} provides targeted information intended to remove the suspected obstacle.
    \item \textbf{Placebo:} provides a structurally matched message without task-relevant guidance.
\end{itemize}

Comparing rescue and placebo allows us to test whether the targeted information affects the observed failure rather than relying only on trajectory inspection.

We study this distinction most closely through Controlled State Reuse (CSR), a task in which an agent must observe a generated state, complete a sequence of intermediate actions, and later reuse that state to obtain the final result. From end-to-end success alone, Gemini 2.5 Flash appears to struggle with state reuse. Checkpoint analysis instead shows that many failures occur earlier, before the model ever observes the state it is expected to reuse. Once that state is observed, downstream completion is generally reliable. A targeted protocol-disambiguation intervention further increases state observation relative to placebo, providing controlled evidence that upstream discovery contributes to the apparent state-reuse failures.

We additionally test whether this diagnosis transfers across model generations by repeating the CSR intervention study with Gemini 3.7 Flash under the same pre-specified experimental design. This cross-generation replication allows us to examine whether a failure mechanism identified for one model remains informative for a newer model without adapting the intervention after observing its behavior.

To make these comparisons reproducible, we pair the diagnostic methodology with a frozen experiment pipeline that preserves the approved task configuration, intervention, execution order, and analysis plan while recording infrastructure failures separately from scientific task outcomes.

In summary, this work makes the following contributions:

\begin{itemize}
    \item We introduce a checkpoint-based methodology for distinguishing failures that occur before a target capability becomes exercisable from failures that occur after exposure, and apply it across four long-horizon security tasks.

    \item We use matched rescue--placebo interventions to test candidate failure mechanisms identified through checkpoint and trajectory analysis.

    \item We show that a discovery bottleneck identified for Gemini 2.5 Flash does not transfer to Gemini 3.7 Flash: the same intervention reverses direction, while substantial failure shifts downstream of state observation.

    \item We develop a reproducible experiment pipeline that preserves the approved experimental design and keeps infrastructure failures separate from scientific outcomes.
\end{itemize}

\section{Background and Related Work}
\label{sec:related}

\subsection{LLM Agents for Offensive Security}

LLMs have increasingly been incorporated into systems that perform multi-step offensive-security workflows. PentestGPT applies LLMs to penetration testing and highlights the challenge of maintaining context across an engagement~\cite{deng2024pentestgpt}. Happe et al. study autonomous Linux privilege escalation and examine the effects of context, guidance, and memory management~\cite{happe2026llms}. Other work demonstrates autonomous website exploitation and exploitation of real-world one-day vulnerabilities~\cite{fang2024hackwebsites,fang2024oneday}, as well as multi-agent approaches for zero-day exploitation~\cite{zhu2024zeroday}. AutoAttacker and PentestAgent further extend agent-based automation to post-breach activity and multi-stage penetration testing~\cite{xu2024autoattacker,shen2024pentestagent}.

More recent work explores agent architectures designed for longer and more complex security workflows. EnIGMA equips a software-engineering agent with interactive security tools~\cite{abramovich2025enigma}, while D-CIPHER separates high-level planning from execution using multiple cooperating agents~\cite{udeshi2025dcipher}. Shell or Nothing combines a realistic penetration-testing benchmark with a memory-activated agent for extended workflows~\cite{mai2025shell}. Bouchari et al. evaluate multiple agent architectures on CTF tasks and study recurring failure patterns and run-to-run consistency~\cite{bouchari2026secondlook}.

These systems show that offensive-security performance depends on more than the underlying language model. The surrounding agent design, including its tools, memory, planning, and interaction interface, can affect performance over extended workflows. Our work does not propose a new offensive-security agent. Instead, we study how to diagnose failures when a security task requires multiple dependent capabilities over a long trajectory.

\subsection{Cybersecurity Agent Benchmarks}

A growing body of work evaluates LLM agents in interactive cybersecurity environments. InterCode provides a general framework for interactive code execution that includes a CTF environment~\cite{yang2023intercode}, while the NYU CTF dataset provides a scalable collection of offensive-security challenges with automated tool-using evaluation~\cite{shao2024nyuctf}. Cybench evaluates agents on professional CTF tasks and includes human-defined subtasks~\cite{zhang2025cybench}. AutoPenBench uses task-specific milestones to evaluate autonomous penetration testing~\cite{gioacchini2025autopenbench}, while PentestEval divides penetration testing into stages and measures performance within each stage~\cite{yang2025pentesteval}.

Other benchmarks move toward real software and more realistic attack environments. CVE-Bench evaluates agents on real-world web vulnerabilities~\cite{zhu2025cvebench}, while CyberGym scales evaluation to vulnerabilities from open-source projects~\cite{wang2025cybergym}. Shell or Nothing evaluates autonomous penetration testing on realistic hosts and services~\cite{mai2025shell}, and AgentCyberRange extends this setting to multi-host environments requiring discovery, exploitation, and post-exploitation~\cite{liu2026agentcyberrange}. CYBERSECEVAL 3 and CAIBench provide broader evaluations spanning multiple cybersecurity capabilities and risks~\cite{wan2024cyberseceval3,sanzgomez2025caibench}.

Several benchmarks also provide information beyond binary task completion. AutoPenBench records task-specific milestones~\cite{gioacchini2025autopenbench}, PentestEval reports performance across different penetration-testing stages~\cite{yang2025pentesteval}, and ExploitBench decomposes exploitation into progressively stronger, independently measurable capabilities~\cite{lee2026exploitbench}. These approaches make it possible to see more of an agent's progress than a single final success rate.

Our work builds on this direction but focuses on a different diagnostic question. We aim to examine whether an agent reaches the point where the capability a task is intended to measure can actually be exercised. This allows us to distinguish failure to reach that point from failure after the capability becomes relevant. We then use matched rescue--placebo interventions to test specific upstream explanations for missed checkpoints.

%Overall, existing work can tell us what progress was made or characterize why trajectories failed. Our specific contribution asks whether the capability under evaluation was ever exercisable, then uses intervention to test a hypothesized upstream cause.

\subsection{Long-Horizon and Process-Level Agent Evaluation}

The challenge of interpreting long interactive trajectories extends beyond cybersecurity. AgentBench and WebArena evaluate agents through repeated interactions with external environments~\cite{liu2023agentbench,zhou2023webarena}, while AppWorld and OSWorld provide executable environments for complex application and computer-use tasks~\cite{trivedi2024appworld,xie2024osworld}. $\tau$-bench evaluates tool-using agents in stateful interactions and emphasizes reliability across repeated trials~\cite{yao2024taubench}. Complementary work studies how agent capability changes with task duration and shows that reliability becomes increasingly important as task horizons grow~\cite{kwa2025longtasks}.

Process-level evaluation provides more information about these trajectories than terminal success alone. AgentQuest measures both progress toward the goal and repeated behavior~\cite{gioacchini2024agentquest}, while AgentBoard tracks incremental progress during multi-turn tasks~\cite{ma2024agentboard}. AgentAtlas analyzes behavioral decisions and trajectory failures~\cite{mazaheri2026agentatlas}, and Odysseys uses graded rubrics to measure partial completion of long-horizon web workflows~\cite{jang2026odysseys}. HORIZON directly studies where and why agents fail as task horizons increase using structured trajectory-level failure attribution~\cite{wang2026horizon}.

Our approach complements these forms of process-level evaluation. Progress metrics, capability ladders, and rubrics describe what an agent accomplishes along a trajectory, while trajectory-level analyses can characterize how and why a run fails. We focus on a narrower question: whether the agent reaches the state required to exercise the capability the task is intended to measure. We then examine behavior after that exposure and, when a specific upstream obstacle can be isolated, test it using a controlled intervention. This prevents failures that occur before capability exposure from being interpreted as failures of the capability itself.

\section{Methodology}
\label{sec:methodology}

Our goal is to diagnose why long-horizon security agents fail. End-to-end success alone cannot distinguish an agent that lacks a capability from one that fails before the capability can ever be exercised. We therefore evaluate agents using intermediate checkpoints that show where progress stops and targeted interventions that test candidate failure mechanisms.

\begin{figure*}[!t]
    \centering
    \includegraphics[width=0.9\textwidth]{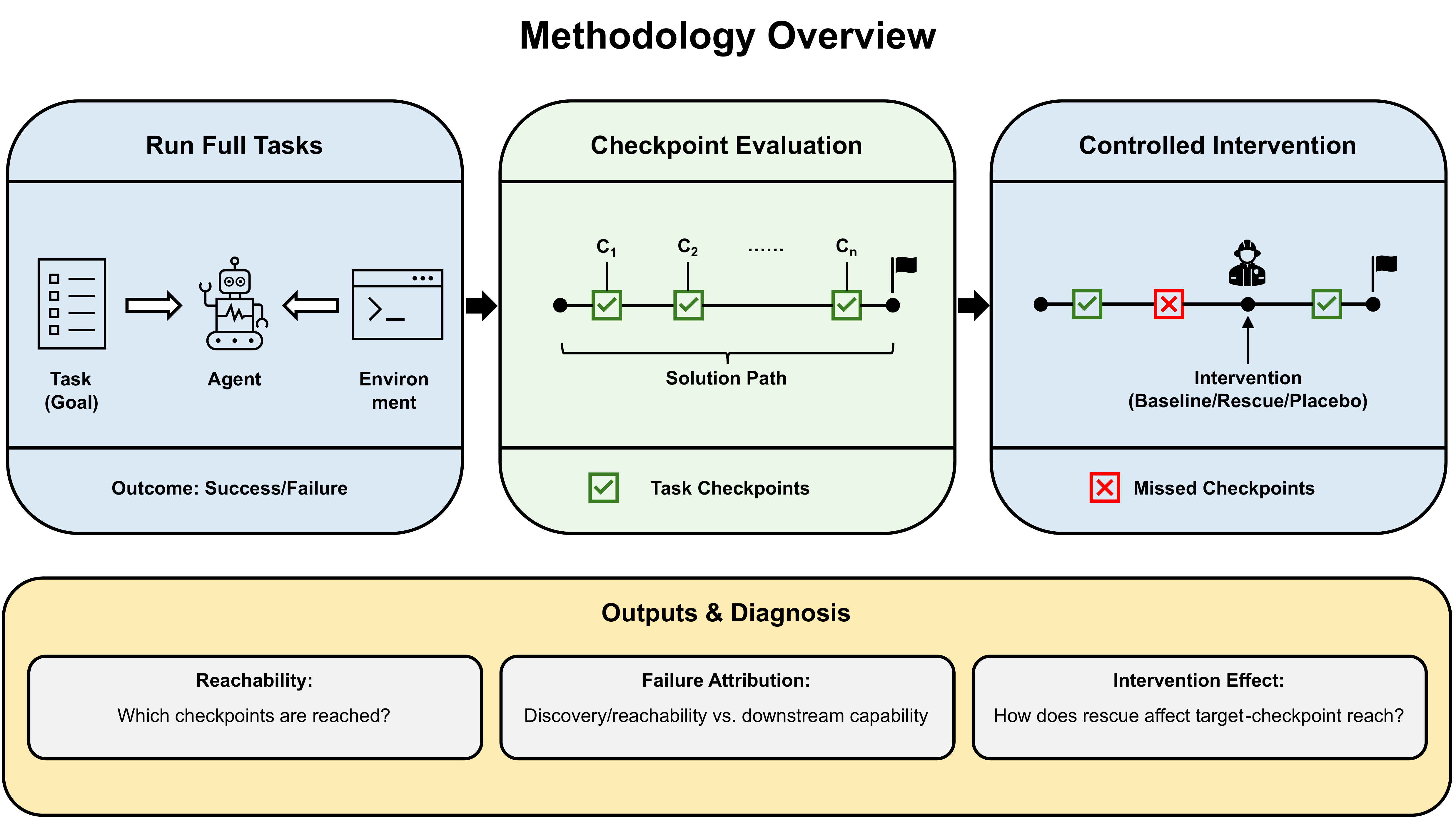}
    \caption{Overview of the diagnostic methodology.}
    \label{fig:framework}
\end{figure*}

Figure~\ref{fig:framework} summarizes the methodology. Each diagnostic task runs in a deterministic, seeded multi-service environment with a fixed agent prompt, task specification, and checkpoint definition. During execution, the evaluator records both final task success and the capability-relevant checkpoints reached by the agent. We use these checkpoints to distinguish \emph{exposure}---whether the agent reaches the point where the target capability can first be exercised---from behavior after that point.

When checkpoint analysis suggests a specific upstream obstacle, we test that explanation using a controlled intervention. Matched runs use the same task instance, model, scaffold, and execution settings, but differ in whether the agent receives targeted information, a structurally matched placebo message, or no message. This allows us to test a candidate failure mechanism experimentally rather than infer it only from trajectory inspection.

We also use an experiment-governance layer to prevent configuration changes or incomplete executions from affecting formal comparisons. Before execution, formal studies fix the run order, relevant configurations, intervention content, source revision, and analysis plan. An append-only experiment record tracks each run and keeps infrastructure failures separate from scientific task outcomes.

\subsection{Diagnostic Task Design}
\label{sec:task-design}

We construct diagnostic tasks around specific long-horizon failure mechanisms. Each task identifies the behavior we want to study and the parts of the task that should remain unchanged across variants. Whenever possible, the environment, discovery process, final interface, agent prompt, and tool protocol remain fixed so that differences between variants reflect the mechanism being studied.

Each task instance has four main components. A deterministic seeded generator first creates task-specific state and service configuration. A Dockerized multi-service environment then exposes the interfaces the agent can interact with, including HTTP services, files, APIs, and SSH. A fixed prompt and tool protocol define how the agent interacts with the environment, while a checkpoint specification maps observable trajectory evidence to task-relevant progress. Before model evaluation, we also verify a valid solution for every task to ensure that the intended workflow is feasible under the frozen configuration.

We instantiate this methodology across four diagnostic tasks spanning three types of long-horizon failure:

\begin{itemize}
    \item \textbf{Delayed Secret Reuse (DSR)} studies whether an agent can retain and later reuse a configuration secret acquired earlier in a multi-stage workflow.
    \item \textbf{Controlled State Reuse (CSR)} studies delayed state reuse while holding discovery and final access fixed and varying only the number of mandatory intermediate actions between state observation and reuse.
    \item \textbf{Strategy Recovery (SR)} studies whether an agent abandons an invalid preferred strategy and transitions to a viable fallback.
    \item \textbf{Ambiguous-Outcome Adaptation (AOA)} studies how an agent responds to an ambiguous preferred-path outcome, including whether it verifies, persists, switches strategies, or abandons progress.
\end{itemize}

% These paper-facing names correspond internally to \texttt{MEM-1B}, \texttt{MEM-1C}, \texttt{LOOP-1}, and \texttt{LOOP-2}, respectively. We retain the original identifiers in the implementation and experiment artifacts so that results reported in the paper can be traced directly to the corresponding code and data.

CSR is our primary controlled task. Its Short, Medium, and Long variants require 2, 8, and 20 mandatory actions, respectively, between observing an exact state and later reusing it. This action count is the only intended difference among the variants, allowing us to separate failure to reach the relevant state from failure to retain and reuse it afterward. DSR is an earlier retention-oriented task, but its Long variant also introduces an additional discovery requirement. We therefore treat DSR as secondary evidence and do not interpret differences between its Short and Long variants as a clean measure of retention.

SR and AOA extend the methodology beyond state reuse. SR studies recovery when a preferred strategy becomes unavailable under control, explicit-failure, and ambiguous-failure conditions. AOA studies behavior after ambiguous outcomes under transient, persistent, and explicit conditions. Together, the four tasks allow us to apply the same diagnostic approach to several forms of long-horizon failure.

\subsection{Checkpoint-Conditioned Exposure and Capability Analysis}
\label{sec:exposure-capability}

Each task contains checkpoints that represent meaningful evidence of progress. A checkpoint is credited only when the required evidence appears in the trajectory, such as in command output or a service response; attempting the corresponding action is not sufficient. When a task uses randomized values, the evaluator may compare the visible evidence against the generated ground truth, but the evidence itself must have been available to the agent. Checkpoints represent task-relevant evidence and do not necessarily form a strictly monotonic sequence. Table~\ref{tab:csr-checkpoints} summarizes the checkpoint sequence used to evaluate CSR trajectories.

\begin{table}[t]
\centering
\caption{CSR checkpoint semantics.}
\label{tab:csr-checkpoints}
\begin{tabular}{cl}
\toprule
Checkpoint & Evidence represented \\
\midrule
C1 & Initial service discovery \\
C2 & Configuration-artifact discovery \\
C3 & Exact generated-state observation \\
C4 & Downstream-interface knowledge \\
C5--C6 & Intermediate-sequence progress \\
C7 & Intermediate sequence completed \\
C8 & Earlier state correctly reused \\
C9 & Final access obtained \\
C10 & Exact final result observed \\
\bottomrule
\end{tabular}
\end{table}

For each capability under study, we identify a \emph{gating checkpoint}: the earliest point at which the agent has the information or state needed to exercise that capability. We refer to reaching this point as \emph{exposure}. We report end-to-end success together with gating-checkpoint reach and success among runs that reach exposure. If many runs fail before exposure but most exposed runs succeed, the observed failures are concentrated upstream of the intended capability.

For Controlled State Reuse, C3 is the gating checkpoint because it records exact observation of the generated state that must later be retained and reused. A run that never reaches C3 never acquires the state required to exercise the reuse capability. We therefore report overall task success, C3 reach, and task success among runs that reach C3.

The same idea applies to other tasks. In Ambiguous-Outcome Adaptation, C5 records the first valid preferred-path outcome and marks the point where the later adaptation decision becomes observable. Runs that terminate before C5 still provide information about overall task reliability, but they do not directly test how the agent responds after observing the relevant outcome.

Exposure conditioning is diagnostic rather than corrective. Runs that fail before exposure remain part of the end-to-end results, and we do not modify the task to force agents to reach the gating checkpoint. Instead, we report overall and exposure-conditioned outcomes together so that failures can be interpreted relative to the capability the task was designed to measure.

Checkpoint analysis shows where failures occur, and we use trajectory inspection to identify candidate explanations and controlled interventions to test those explanations when the suspected obstacle can be isolated.

\subsection{Controlled Interventions for Causal Diagnosis}
\label{sec:interventions}

A controlled intervention targets one hypothesized upstream obstacle while leaving the rest of the task unchanged. We then test whether removing or reducing that obstacle changes the probability of reaching the relevant checkpoint. Final task success is also measured, but is secondary because it depends on later stages of the workflow.

Interventions can be triggered at a fixed model turn or by a specified execution event. For example, an event-based intervention can be delivered after a predefined occurrence of \texttt{command\_completed} and immediately before the next model call. The trigger does not depend on whether the task is succeeding and does not inspect generated task values to decide when to intervene. We separately record whether a run reaches the intervention point and whether the intervention is delivered.

We evaluate interventions using matched triples. For every deterministic environment seed, we run baseline, rescue, and placebo conditions using the same model, task instance, scaffold, execution limits, and intervention timing. Baseline receives no additional message. Rescue receives information targeted at the suspected obstacle, while placebo receives a structurally matched message without task-relevant guidance. Rescue versus placebo is the primary comparison because it isolates the targeted information from the general effect of inserting an additional message. Baseline provides descriptive context.

For the CSR discovery intervention, rescue provides only the protocol information needed to distinguish the HTTP services from the SSH service. It does not disclose the configuration artifact, the generated state, the downstream interface, any credentials, or the final outcome. The agent must still locate the artifact, inspect its contents, complete the remaining steps, maintain the generated state throughout execution, and reuse it when necessary. As a result, the intervention isolates the suspected discovery bottleneck without materially simplifying the rest of the task.

Rescue and placebo messages are matched in byte and line count and frozen before formal execution. The analysis record stores the experimental condition, intervention category, whether the intervention point was reached, and whether the message was delivered. Task-specific intervention content is not copied into the analysis metadata.

The primary outcome for each intervention is the checkpoint it is intended to affect. For the CSR discovery study, the primary outcome is C3 reach. End-to-end success is secondary because it also depends on later task stages. Because rescue and placebo runs use the same environment seed, their binary outcomes are compared using a two-sided exact McNemar test~\cite{mcnemar1947sampling}.

Trajectory patterns are therefore treated as hypotheses rather than conclusions. An intervention may support the suspected mechanism, produce no detectable effect, or affect behavior in the opposite direction. All three outcomes remain informative and are interpreted under the same frozen study design.

We also test whether an intervention effect transfers across model generations. For Gemini 3.7 Flash, we repeat the CSR discovery study using the same CSR-Medium task, environment seeds, seed-condition execution order, intervention trigger and messages, agent scaffold, execution limits, checkpoint definitions, primary outcome, and statistical test used for the Gemini 2.5 Flash confirmatory study. This replication is pre-specified before observing Gemini 3.7 outcomes, while direct cross-generation comparisons are treated as secondary and descriptive.

\subsection{Experimental Governance and Reproducibility}
\label{sec:governance}

Long-running agent experiments can be affected by changes in task generation, intervention content, source code, execution order, or recovery from interrupted runs. These changes are especially important in confirmatory intervention studies because modifications made after observing results could alter the comparison. We therefore freeze the approved experimental state before formal execution and record the lifecycle of every planned run.

Each experiment is defined by a manifest containing the planned runs and their task variant, environment seed, model, provider settings, execution limits, condition assignment, and intervention trigger. The execution order is generated and stored before the experiment begins. During execution, the runner selects the next eligible position from this fixed order and the existing experiment record rather than generating or choosing runs dynamically.

Important experimental inputs are protected using cryptographic digests. These include the scaffold configuration and, for intervention studies, the rescue and placebo messages. Intervention messages are also checked against structural requirements such as byte and line count. If bound content changes, validation fails rather than allowing the modified content to be executed under the same experiment identifier.

Formal studies follow a candidate, review, and approval process. Before approval, the candidate experiment is checked for matched-set consistency, intervention-trigger consistency, content bindings, and execution-order integrity. After review, the finalized manifest and source revision are recorded in a separate approval artifact. The runner verifies these bindings before allowing formal execution.

Approved studies are also tied to the source revision used by the evaluator, runner, environment, and experiment infrastructure. Once formal execution begins, both the approved manifest and the relevant source state are treated as frozen.

Run state is stored in an append-only ledger. Each planned run receives lifecycle records for preparation, execution, completion or failure, and evaluation. Existing records are not overwritten to turn interrupted runs into successful ones, and the same ledger allows interrupted batches to resume without changing the predetermined execution order.

We use infrastructure failure to refer to failures in our experimental execution stack, such as runner or executor errors that prevent a run from producing a valid measurement. These are distinct from task failures, where the agent executes normally but does not complete the security task. Infrastructure failures are recorded separately from task failures. Such runs remain in the experiment record and are not silently rerun or replaced, allowing the analysis to distinguish invalid measurements from valid task failures.

% These mechanisms make the experiment reproducible beyond fixing a random seed. They preserve the task definition, source state, intervention content, execution order, and approved study design while keeping operational failures visible in the experimental record.

\subsection{Statistical Analysis}
\label{sec:statistics}

For each task and model, we report end-to-end success, reach of the relevant gating checkpoint, and success among runs that reach that checkpoint. These results are used to determine where failures occur, rather than to make population-level claims from the small exploratory cohorts.

Controlled intervention studies use matched environment seeds. Each seed contributes baseline, rescue, and placebo runs under the same frozen task and execution settings. Rescue versus placebo is the primary comparison because it tests the effect of the targeted information, while baseline comparisons provide descriptive context.

We use the initial five-seed CSR study to finalize the intervention and plan the sample size for the Gemini 2.5 Flash confirmatory study, but do not pool its outcomes with the confirmatory results. The confirmatory study uses 92 additional matched seeds. Because rescue and placebo runs are paired by seed, sample-size planning is based on the expected number and direction of rescue--placebo disagreements. We use paired power analysis for McNemar's test, checked using both a normal approximation and exact enumeration; an independent-groups calculation serves only as a conservative cross-check. This analysis yields the 92-seed Gemini 2.5 design, which is reused unchanged for the Gemini 3.7 replication rather than recalculated for the newer model.

For the Gemini 2.5 Flash confirmatory study, the primary outcome is C3 reach under rescue versus placebo. We compare these paired binary outcomes using a two-sided exact McNemar test with $\alpha=0.05$ and report both marginal reach rates and the two directions of discordant pairs. End-to-end success is analyzed with the same paired test as a secondary outcome. Baseline comparisons and success conditioned on C3 are descriptive.

McNemar's test considers only matched seeds where rescue and placebo produce different outcomes. In our setting, these are seeds where rescue reaches C3 and placebo does not, or where placebo reaches C3 and rescue does not. Under the null hypothesis that the intervention has no effect, the two directions of disagreement are equally likely. The exact two-sided $p$-value measures how unlikely an imbalance at least as large as the observed one would be under this null hypothesis. We use $\alpha=0.05$, so $p<0.05$ is treated as evidence against the null hypothesis.

The Gemini 3.7 Flash replication uses the same 92 environment seeds, task, intervention trigger and messages, execution order, checkpoint definitions, primary outcome, and exact McNemar test. Its analysis plan is frozen before any Gemini 3.7 outcomes are examined. Rescue versus placebo is analyzed as a separate confirmatory test for Gemini 3.7, while comparisons between the two model generations remain secondary and descriptive.

Infrastructure failures are reported separately from task failures and are not replaced by reruns. The original Gemini 2.5 pre-specification required runs to reach the intervention point but did not explicitly address truncated infrastructure-invalid executions. Before examining C3 or success outcomes, we documented an analysis clarification excluding such runs from the paired analysis and report sensitivity bounds for the affected missing placebo observations. The Gemini 3.7 pre-specification includes this infrastructure-validity rule explicitly before execution.

For Gemini 3.7, a rescue--placebo pair enters the primary analysis only when both runs are infrastructure-valid and reach the intervention point. We report excluded pairs and their reasons and do not impute missing scientific outcomes. When exclusions affect the primary pair set, we also report extreme sensitivity bounds separately from the pre-specified primary analysis.

For both confirmatory studies, the primary outcome, paired test, significance threshold, and distinction between primary and secondary analyses are fixed before the corresponding outcomes are examined. The exploratory pilot, Gemini 2.5 Flash confirmatory study, and Gemini 3.7 Flash replication are therefore kept separate throughout the analysis.

\section{Experimental Setup}
\label{sec:setup}

We evaluate our methodology on four long-horizon security tasks implemented in deterministic Docker environments. Each task requires the agent to interact with multiple services over several dependent actions. Task-specific values are generated from a deterministic seed, allowing matched runs to use the same underlying task instance.

\subsection{Tasks and Environments}
\label{sec:setup-tasks}

\begin{figure*}[ht!]
    \centering
    \includegraphics[width=0.9\textwidth]{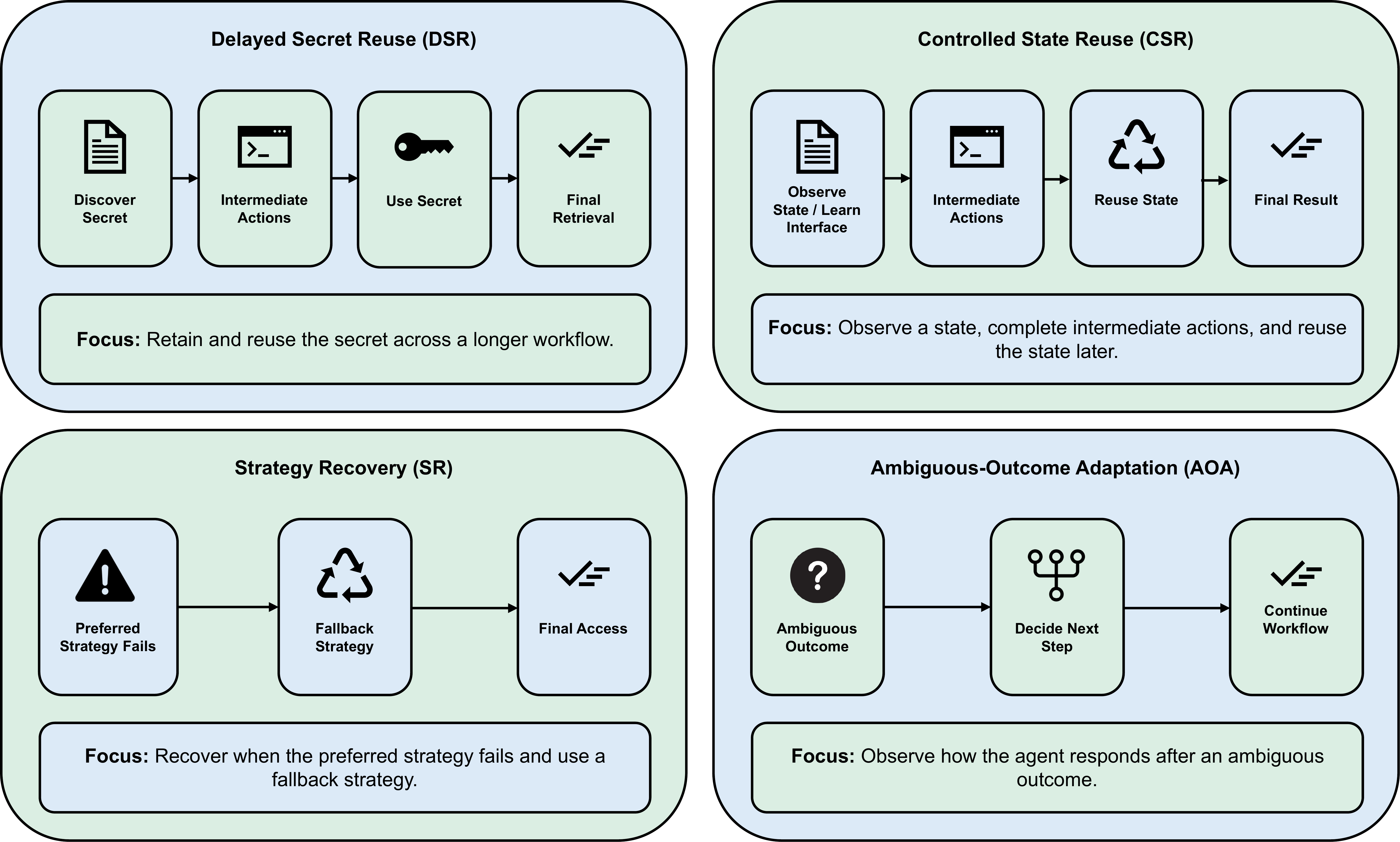}
    \caption{Overview of the four diagnostic task families.}
    \label{fig:task-families}
\end{figure*}

Table~\ref{tab:tasks} summarizes and Figure~\ref{fig:task-families} shows the overview of the four task families.

\begin{table*}[!t]
\centering
\caption{Diagnostic tasks used in our evaluation.}
\label{tab:tasks}
\begin{tabular}{lll}
\toprule
Task & Variants & Primary focus \\
\midrule
Delayed Secret Reuse (DSR) & Short, Long & Delayed secret reuse \\
Controlled State Reuse (CSR) & Short, Medium, Long & Controlled state reuse \\
Strategy Recovery (SR) & Control, Explicit, Ambiguous & Recovery after failure \\
Ambiguous-Outcome Adaptation (AOA) & Transient, Persistent, Explicit & Decision under ambiguity \\
\bottomrule
\end{tabular}
\end{table*}

The environments expose services and resources through HTTP, files, APIs, and SSH. Secrets, configuration values, and other task-specific state are generated deterministically from the environment seed. Within matched comparisons, the prompt and tool interface remain unchanged, while the task specification determines the intended difference between variants.

\textbf{Delayed Secret Reuse (DSR).} DSR requires the agent to discover a configuration secret early in the workflow, reuse it later at a protected API endpoint, and then complete an SSH-gated final retrieval. The Short and Long variants were intended to vary the gap between acquiring and reusing the secret. We use DSR only as secondary evidence since the Long variant changes both the delay before reuse and the route-discovery requirement, making it difficult to isolate retention as the only source of difference.

\textbf{Controlled State Reuse (CSR).} CSR provides a more controlled test of delayed state reuse. The agent must discover an exact generated state, learn the downstream interface, complete a required sequence of intermediate actions, and later reuse the earlier state to obtain the final result. CSR-Short, CSR-Medium, and CSR-Long require 2, 8, and 20 intermediate actions, respectively, and this action count is the only intended difference among the variants. C3 records exact observation of the generated state and is the exposure checkpoint used in our primary analysis.

\textbf{Strategy Recovery (SR).} SR tests whether the agent can recover when an initially preferred strategy becomes unavailable. All variants provide the same final-access interface and the same viable fallback, which requires six ordered actions. In the Control variant, the preferred strategy succeeds. The Explicit and Ambiguous variants instead block it with either a clear terminal failure or an ambiguous failure. The fallback path remains unchanged.

\textbf{Ambiguous-Outcome Adaptation (AOA).} AOA studies how the agent responds after an ambiguous outcome on its preferred path. If the preferred strategy is revisited, the Transient variant succeeds after the intended verification retry, the Persistent variant remains ambiguous, and the Explicit variant returns a terminal failure. These variants let us study how agents adapt when the outcome of a preferred strategy is uncertain.

CSR, SR, and AOA use a common C0--C10 checkpoint numbering scheme with task-specific checkpoint semantics. DSR predates this scheme and retains its own task-specific checkpoints and failure labels.

\subsection{Models and Agent Execution}
\label{sec:setup-models}

Our initial cross-model evaluation uses Gemini 2.5 Flash and Gemini 2.5 Pro. Within matched comparisons, both models use the same agent scaffold, tool protocol, task-specific prompt, and execution limits.

CSR and AOA form the main cross-model cohorts. We evaluate both Gemini 2.5 models on CSR-Short, CSR-Medium, and CSR-Long using matched environment seeds. We also evaluate both models on the Transient and Persistent AOA variants used for exposure analysis. SR includes both models across all three variants, while DSR is evaluated only with Gemini 2.5 Flash.

Our CSR analysis proceeds in stages. We first use the initial Gemini 2.5 CSR runs to characterize where failures occur and identify protocol discovery as a candidate upstream obstacle for Gemini 2.5 Flash. We then evaluate a rescue--placebo intervention targeting this obstacle in a small exploratory study and use that study to finalize the intervention design. The finalized design is tested in an independent confirmatory study on CSR-Medium. Finally, we repeat the same confirmatory design with Gemini 3.7 Flash to examine whether the diagnosed failure pattern and intervention effect transfer to a newer model generation.

The controlled CSR studies use a fixed tool-action protocol and stopping policy. Each run is limited to 90 executed commands and 120 model turns, and terminates after 12 consecutive protocol errors. Commands time out after 30\,s and provider requests after 60\,s. Model responses are limited to 1,024 output tokens, with at most two retries for provider calls. These limits are identical across baseline, rescue, and placebo runs and are preserved in the Gemini 3.7 Flash replication.

All model calls are made through OpenRouter. The same experiment runner and evaluator are used across individual and batch executions so that task selection, intervention delivery, run recording, and checkpoint evaluation follow the same procedure.

\subsection{Evaluation Protocol}
\label{sec:setup-protocol}

Each run is evaluated for both final task completion and checkpoint evidence. Our evaluation proceeds in three stages: initial characterization, exploratory intervention testing, and confirmatory intervention testing. The initial characterization is descriptive and is used to identify where failures occur and to generate candidate explanations; formal statistical testing is reserved for the confirmatory intervention studies.

For CSR, each Gemini 2.5 model is first evaluated in 15 runs: five seeds each for the Short, Medium, and Long variants. These runs reveal the initial exposure pattern and, through trajectory inspection, motivate protocol discovery as a candidate upstream obstacle for Gemini 2.5 Flash. The AOA exposure cohort contains 10 runs per model across the Transient and Persistent variants. SR is evaluated across all three variants with both Gemini 2.5 models, while DSR uses five Short and five Long runs with Gemini 2.5 Flash.

We next evaluate the proposed CSR discovery intervention using five matched environment seeds. Each seed contributes one baseline, one rescue, and one placebo run, giving 15 exploratory Gemini 2.5 Flash runs in total. Baseline, rescue, and placebo use the same task instance, model, scaffold, execution limits, and intervention timing. Rescue and placebo differ only in message content, while baseline receives no intervention. We use this exploratory study to finalize the intervention design and determine the confirmatory sample size; its seeds and outcomes are kept separate from the confirmatory analysis.

The confirmatory Gemini 2.5 Flash study then uses 92 new matched seeds, producing 276 planned runs: 92 baseline, 92 rescue, and 92 placebo. This study tests the finalized intervention independently of the exploratory runs.

In the confirmatory study, rescue and placebo messages are delivered after the second completed command. Rescue provides only the protocol mapping needed to distinguish HTTP services from the SSH service. Placebo provides no task-relevant protocol information. Neither message reveals the configuration artifact, generated state, downstream route, credentials, or final result.

The Gemini 3.7 Flash replication uses the same 92 environment seeds and preserves the seed-condition execution order from the Gemini 2.5 study. It also keeps the CSR-Medium task, intervention trigger and messages, agent scaffold, execution limits, and checkpoint definitions unchanged. The replication is separately pre-specified and approved before execution.

All formal runs use the checkpoint definitions and completion rules fixed for the corresponding approved study. Infrastructure failures are recorded separately from task failures and remain in the experiment record rather than being silently rerun or replaced.

\section{Evaluation}
\label{sec:evaluation}

We first use checkpoint evidence to determine whether end-to-end failures occur before or after the capability each task is intended to measure becomes exercisable. We then test the discovery bottleneck identified for Gemini 2.5 Flash on CSR using a controlled intervention and repeat the same frozen study with Gemini 3.7 Flash. Finally, we report secondary findings from DSR, SR, and AOA.

\subsection{Exposure Before Capability}
\label{sec:eval-exposure}

\begin{figure*}[!t]
    \centering
    \includegraphics[width=0.9\textwidth]{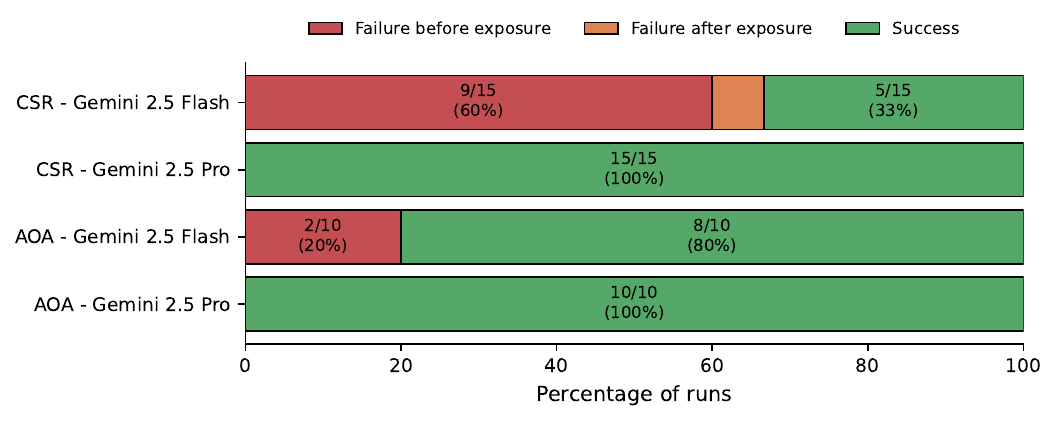}
    \caption{Exposure-conditioned outcome decomposition.}
    \label{fig:exposure-decomposition}
\end{figure*}

Figure~\ref{fig:exposure-decomposition} separates each run into success, failure before exposure, or failure after exposure. For Gemini 2.5 Flash, much of the observed performance gap occurs before the capability-relevant checkpoint is reached.

The clearest example is CSR. Gemini 2.5 Flash succeeds in 5/15 runs, compared with 15/15 for Gemini 2.5 Pro. However, Flash reaches C3, where it first observes the state required for later reuse, in only 6/15 runs. Of those six runs, five succeed. Pro reaches C3 and succeeds in all 15 runs. Thus, nine of the ten failed Flash runs never acquire the state needed to exercise the reuse capability. Its 33\% end-to-end success rate therefore reflects mostly failure before exposure rather than failure to reuse an observed state.

AOA shows a similar pattern. Gemini 2.5 Flash succeeds in 8/10 runs and reaches the C5 decision point in exactly those eight runs; every Flash run that reaches C5 subsequently succeeds. Gemini 2.5 Pro reaches C5 and succeeds in all 10 runs. Although AOA studies a different capability, its lower Flash success rate also originates before the capability-relevant checkpoint.

Flash's C3 reach also varies across the three CSR variants: 1/5 for Short, 1/5 for Medium, and 4/5 for Long, while Pro reaches C3 in 5/5 runs for every variant. We treat these small-sample differences descriptively. In particular, they should not be interpreted as evidence that a longer intermediate sequence improves discovery, because C3 occurs before the controlled intermediate sequence begins.

Trajectory inspection suggested one possible reason for the missed C3 checkpoint. In a non-reaching Flash run, the agent repeatedly attempted unauthenticated SSH access instead of performing the HTTP reconnaissance needed to discover the relevant state. Similar behavior appeared in other non-reaching CSR runs. We treat this pattern as a hypothesis rather than an explanation and illustrate it with a matched case study next.

\begin{mdframed}[backgroundcolor=gray!10]
\noindent
\textbf{Key Finding:}
Most Gemini 2.5 Flash failures in CSR occur before the state required for reuse is observed. Trajectory inspection suggests protocol discovery as a possible upstream obstacle.
\end{mdframed}

\subsection{Case Study: From Discovery Failure to Successful Completion}
\label{sec:eval-case-study}

To illustrate the failure pattern targeted by our intervention, we examine one matched CSR-Medium seed in which the placebo run fails while the rescue run succeeds. Both runs use the same generated environment, model, scaffold, and execution limits; they differ in the content of the intervention message.

In the placebo run, the agent repeatedly attempts to access the SSH service without first performing the HTTP reconnaissance needed to discover the configuration artifact and generated state. As a result, the run never reaches the state-observation checkpoint and cannot exercise the later state-reuse capability. From the end-to-end outcome alone, this appears to be a state-reuse failure even though the required state was never observed.

In the matched rescue run, the agent receives only the protocol mapping that distinguishes the HTTP services from the SSH service. The intervention does not reveal the configuration artifact, generated state, downstream interface, credentials, or final result. With this information, the agent follows the appropriate discovery path, observes the generated state, completes the remaining workflow, and ultimately succeeds.

This example shows why exposure matters for interpreting task failure. The placebo run fails before state reuse can be exercised, while the matched rescue run reaches the required state and completes the task after receiving targeted protocol information. A single pair cannot establish that protocol discovery is the general cause of missed exposure, so we test the same hypothesis across the full confirmatory study below.

\begin{mdframed}[backgroundcolor=gray!10]
\noindent
\textbf{Key Finding:}
In this matched example, the placebo run fails before state observation, while protocol-disambiguation guidance enables the rescue run to reach the required state and complete the task. The confirmatory study next tests whether this pattern generalizes.
\end{mdframed}

\subsection{Causal Diagnosis of the CSR Discovery Bottleneck}
\label{sec:eval-causal}

Our first intervention targeted state reuse later in the CSR workflow by restoring only the exact state value that the agent had previously observed. However, only 2 of 5 rescue--placebo pairs reached the intervention point in both conditions, and both pairs succeeded under both rescue and placebo. Because there were no discordant pairs, the experiment provided little information about the effect of restoring state. More importantly, it confirmed that many runs were failing before a reuse-stage intervention could even be applied.

We therefore moved the intervention earlier and targeted the protocol-selection behavior suggested by trajectory inspection. In a five-seed exploratory study, C3 was reached in 5/5 rescue runs, 4/5 placebo runs, and 2/5 baseline runs. We use it to finalize the intervention and evaluate the same design in an independent confirmatory study.

The Gemini 2.5 Flash confirmatory study contains 276 planned runs, all of which reached a terminal experiment state. Seven runs ended because of the same infrastructure-level executor exception: five placebo, two baseline, and no rescue runs. These runs were recorded but not rerun or replaced. Because the five invalid placebo runs cannot form complete rescue--placebo comparisons, the primary analysis contains 87 matched pairs.

The intervention has a large effect on C3 reach. Rescue reaches C3 in 83/87 runs (95.4\%), compared with 57/87 (65.5\%) under placebo, a difference of 29.9 percentage points. Among the 32 pairs in which rescue and placebo differ, 29 favor rescue and 3 favor placebo. The exact two-sided McNemar test gives $p=2.56\times10^{-6}$, and the rescue-favoring fraction among discordant pairs is 90.6\% with an exact 95\% confidence interval of [75.0\%, 98.0\%]. The other 55 pairs are concordant: 54 reach C3 in both conditions and one reaches C3 in neither. These results show that providing only protocol-disambiguation information substantially increases C3 reach relative to a structurally matched task-neutral message.

End-to-end success follows nearly the same pattern. Rescue succeeds in 82/87 runs (94.3\%), compared with 56/87 (64.4\%) under placebo, with 29 rescue-only and 3 placebo-only successes. We treat final success as a secondary outcome because the intervention directly targets discovery, while task completion also depends on later stages of the workflow.

After C3 is reached, Gemini 2.5 Flash usually completes the remaining workflow. Among infrastructure-valid runs across baseline, rescue, and placebo that reach both the intervention point and C3, 205/210 (97.6\%) succeed: 63/65 baseline, 86/88 rescue, and 56/57 placebo. This supports the earlier exposure analysis: in this study, the main limitation is reaching the relevant state, not completing the task after observing it.

Baseline comparisons provide descriptive context for the effect of receiving an additional message. Among 85 seeds with valid baseline and placebo runs, C3 is reached in 62/85 baseline runs (72.9\%) and 55/85 placebo runs (64.7\%). This is opposite to the direction observed in the five-seed pilot, so we do not interpret the pilot's placebo--baseline difference as evidence that message injection itself is beneficial. Among 90 valid rescue--baseline pairs, rescue reaches C3 in 86/90 runs (95.6\%) compared with 65/90 (72.2\%) for baseline.

The infrastructure failures require one analysis clarification. The original pre-specified population rule required runs to reach the intervention point but did not explicitly address executions truncated by infrastructure failure. Before examining C3 or success outcomes, we documented a clarification excluding these truncated runs from the paired analysis because they did not provide normally completed measurements. We report this change explicitly rather than presenting it as part of the original population rule.

Because all five invalid rescue--placebo observations occur on the placebo side, we also test both extreme assumptions for their missing C3 outcomes. If all five are treated as reaching C3, the exact McNemar $p$-value is $2.56\times10^{-6}$; if all five are treated as not reaching C3, it is $1.23\times10^{-7}$. The conclusion is unchanged under either assumption.

These results provide controlled evidence that protocol discovery contributes causally to C3 exposure for Gemini 2.5 Flash on CSR-Medium. We next test whether the same diagnosis and intervention effect remain stable for a newer model generation.

\begin{mdframed}[backgroundcolor=gray!10]
\noindent
\textbf{Key Finding:}
For Gemini 2.5 Flash on CSR-Medium, protocol discovery is a causal contributor to the observed exposure bottleneck. Once the relevant state is reached, downstream task completion is highly reliable.
\end{mdframed}

\subsection{Cross-Generation Replication and Intervention Reversal}
\label{sec:eval-cross-generation}

We repeat the same frozen confirmatory design with Gemini 3.7 Flash without adapting the intervention after observing the newer model's behavior. All 276 planned runs reach a terminal experiment state. Three runs are infrastructure-invalid, but only one placebo run affects rescue--placebo eligibility. The primary Gemini 3.7 analysis therefore contains 91 matched pairs.

\begin{figure}[t]
    \centering
    \includegraphics[width=\columnwidth]{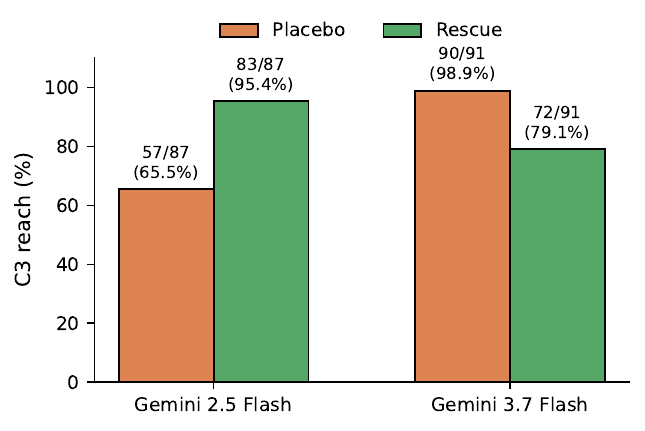}
    \caption{Confirmatory C3 reach across model generations.}
    \label{fig:cross-generation-c3}
\end{figure}

Figure~\ref{fig:cross-generation-c3} shows that the intervention effect reverses across model generations. For Gemini 2.5 Flash, rescue increases C3 reach from 65.5\% to 95.4\%. For Gemini 3.7 Flash, rescue instead reaches C3 in 72/91 runs (79.1\%), compared with 90/91 (98.9\%) under placebo.

Among the 20 discordant Gemini 3.7 pairs, 19 favor placebo and only one favors rescue. The exact two-sided McNemar test gives $p=4.01\times10^{-5}$, with a rescue-minus-placebo difference of $-19.8$ percentage points. Thus, the same protocol-disambiguation intervention that improves C3 reach for Gemini 2.5 Flash produces a statistically significant effect in the opposite direction for Gemini 3.7 Flash.

\begin{table}[t]
\centering
\caption{Cross-generation CSR intervention results.}
\label{tab:cross-generation}
\begin{tabular}{lcc}
\toprule
 & Gemini 2.5 Flash & Gemini 3.7 Flash \\
\midrule
Rescue C3 reach
    & 83/87 (95.4\%)
    & 72/91 (79.1\%) \\
Placebo C3 reach
    & 57/87 (65.5\%)
    & 90/91 (98.9\%) \\
Discordant pairs (R/P)
    & 29 / 3
    & 1 / 19 \\
McNemar $p$
    & $2.56\times10^{-6}$
    & $4.01\times10^{-5}$ \\
Success given C3
    & 205/210 (97.6\%)
    & 90/249 (36.1\%) \\
\bottomrule
\end{tabular}
\end{table}

The reversal is robust to the single excluded placebo observation. If its missing C3 outcome is treated as a failure, the exact McNemar test gives $p=2.21\times10^{-4}$; if it is treated as a success, the result is $p=4.01\times10^{-5}$. Both assumptions preserve the placebo-favoring direction and statistical significance.

Reaching C3 also has a very different relationship with final success for Gemini 3.7 Flash. Across eligible baseline, rescue, and placebo runs, only 90/249 C3-reaching runs (36.1\%) ultimately succeed, compared with 205/210 (97.6\%) for Gemini 2.5 Flash. Gemini 3.7 therefore often observes the exact state required for CSR but still fails later in the workflow. For this model, failure to reach C3 is no longer the main explanation for end-to-end failure.

Final task success moves in the same direction as the C3 result but does not reach statistical significance. Rescue succeeds in 28/91 runs (30.8\%), compared with 40/91 (44.0\%) under placebo. The 40 discordant pairs include 14 rescue-only and 26 placebo-only successes, giving an exact McNemar $p=0.0807$. We therefore retain C3 reach as the pre-specified primary result and treat final task success as secondary.

We also compare intervention direction across the subset of seeds that are eligible in both model studies. Within this common set, Gemini 2.5 Flash has 29 rescue-favoring and 3 placebo-favoring discordant pairs, corresponding to a rescue-favoring fraction of 0.906 (95\% CI [0.750, 0.980]). Gemini 3.7 Flash has 1 rescue-favoring and 18 placebo-favoring pairs, corresponding to 0.053 (95\% CI [0.001, 0.260]). The Gemini 3.7 count differs from the 1/19 primary result above because this analysis uses only the cross-generation intersection of eligible seeds. Because the cross-generation comparison was pre-specified as descriptive, we do not add a post-hoc significance test between models.

The replication changes the diagnostic picture rather than invalidating the Gemini 2.5 result. For Gemini 2.5 Flash, protocol disambiguation substantially improves state observation, and runs that reach the state usually complete the remaining workflow. For Gemini 3.7 Flash, state observation is already common under placebo, the same intervention reduces C3 reach, and substantial failure remains after C3. Thus, protocol discovery is no longer the dominant limitation for the newer model. More generally, a bottleneck identified for one model generation should not be assumed to remain the main source of failure for another, even when the task, scaffold, and intervention are held fixed.

\begin{mdframed}[backgroundcolor=gray!10]
\noindent
\textbf{Key Finding:}
The dominant failure pattern changes across model generations. Protocol discovery is an important exposure bottleneck for Gemini 2.5 Flash, but is no longer the dominant limitation for Gemini 3.7 Flash, where substantial failure remains after the required state is observed.
\end{mdframed}

\subsection{Secondary Diagnostic Results}
\label{sec:eval-secondary}

The remaining tasks show other ways in which checkpoint evidence changes the interpretation of final success.

\textbf{Delayed Secret Reuse.}
Gemini 2.5 Flash succeeds in 4/5 DSR-Short runs and 2/5 DSR-Long runs. At first, the lower Long-variant success rate could appear to show greater difficulty retaining a secret across a longer workflow. However, all five Long runs discover the secret, while only two complete the additional route-discovery stage introduced by that variant; both of those runs later reuse the secret correctly. Because DSR-Long changes both dependency length and route discovery, the result does not provide clean evidence of worse retention.

\textbf{Strategy Recovery.}
SR reaches a performance ceiling: all 30 runs across both Gemini 2.5 models and all three variants succeed. This ceiling is itself diagnostically informative because a task may be valid in design yet provide little value for distinguishing models once evaluated agents solve it consistently. SR therefore illustrates that high task success does not necessarily mean that a benchmark remains informative for comparing the capability of interest.

\textbf{Ambiguous-Outcome Adaptation.}
AOA has the opposite problem: the intended comparison is not exercised. The Transient-versus-Persistent comparison requires the agent to revisit the preferred strategy after an initial ambiguous outcome. However, the C6 checkpoint corresponding to the second ambiguous outcome is reached in 0/20 runs in the Transient--Persistent exposure cohort. Agents generally switch away from the preferred strategy after the first ambiguous signal, so the intended retry-policy comparison is never reached.

AOA still provides useful exposure evidence. Gemini 2.5 Flash fails to reach the initial C5 decision point in 2/10 runs, while Gemini 2.5 Pro reaches it in all 10. Every Flash run that reaches C5 subsequently succeeds. Thus, AOA again shows how checkpoint reach can change the interpretation of aggregate success even when the task does not exercise the comparison it was originally designed to study.

These secondary results show three different limitations that final success alone would hide. DSR changes more than one factor between variants, SR provides no failures to diagnose, and AOA does not reach the branch required for its intended comparison. Checkpoint evidence makes these limitations visible and prevents the corresponding end-to-end results from being overinterpreted.

\begin{mdframed}[backgroundcolor=gray!10]
\noindent
\textbf{Key Finding:}
Checkpoint evidence reveals limitations that final success rates can hide: DSR contains a confounded comparison, SR reaches a performance ceiling, and AOA does not exercise its intended experimental branch.
\end{mdframed}

\section{Discussion}
\label{sec:discussion}

\subsection{Limitations of End-to-End Success Metric}

End-to-end success and capability diagnosis answer different questions. Final success shows whether the full workflow was completed, but not whether the agent ever reached the point where the capability of interest could be exercised. In CSR, this distinction changes the interpretation of Gemini 2.5 Flash: what initially appears to be poor state reuse is largely failure to reach the state that must later be reused.

The secondary tasks show why this distinction matters more broadly. DSR changes more than one factor between variants, SR reaches a performance ceiling, and AOA rarely reaches the branch needed for its intended comparison. In each case, checkpoint evidence limits what can reasonably be concluded from the final success rate.

We therefore view end-to-end and exposure-conditioned results as complementary. End-to-end success measures overall reliability, while checkpoint analysis shows where failures occur relative to the capability being studied. Claims about a specific capability should consider both.

\subsection{From Failure Localization to Causal Diagnosis}

Checkpoint analysis can show where progress stops, but it does not by itself explain the underlying cause. A missed checkpoint may have several plausible explanations, especially in long trajectories where many earlier decisions can affect later progress.

We therefore treat patterns found through trajectory inspection as hypotheses to test. For Gemini 2.5 Flash, inspection suggested that protocol discovery was preventing the agent from reaching C3. Providing only the missing protocol information substantially increased C3 reach relative to a matched placebo. This does not reveal every detail of the model's reasoning, but it provides controlled evidence that the targeted discovery obstacle contributes to the failure.

Checkpoint analysis and controlled intervention therefore serve different roles. Checkpoints localize the failure, while interventions test whether a specific suspected obstacle actually contributes to it.

\subsection{Failure Modes Can Shift Across Model Generations}

The Gemini 3.7 replication shows that a failure mechanism identified for one model should not be assumed to transfer to another. Under the same CSR task and frozen intervention design, the protocol guidance that improves C3 reach for Gemini 2.5 Flash instead reduces it for Gemini 3.7 Flash. Reaching C3 also no longer reliably leads to downstream completion for the newer model.

Gemini 3.7 therefore does not simply show a weaker version of the Gemini 2.5 pattern. For Gemini 2.5, failure is concentrated before state observation. For Gemini 3.7, state observation is common, but substantial failure remains afterward. The same checkpoint can therefore have a different diagnostic meaning for different models.

This matters for comparisons across model generations. A benchmark may remain unchanged while the location and cause of failure shift. Similar final success rates can therefore arise from different underlying failure patterns, and an intervention designed for one model's bottleneck should not automatically be treated as a general solution.

The reversal also shows why preserving the original intervention design is useful. Had we adapted the intervention after observing Gemini 3.7 behavior, we might have missed that the earlier diagnosis no longer transferred. A diagnostic evaluation should be able to reveal not only when a proposed explanation is supported, but also when it stops explaining the behavior of a newer model.

\subsection{Implications for Security-Agent Benchmark Design}

Our findings suggest several practical lessons for long-horizon security-agent benchmarks.

First, checkpoints should correspond to evidence that is actually visible to the agent and should be tied to the capability being evaluated. This makes it possible to distinguish failure to reach the relevant state from failure after reaching it.

Second, task variants should change only the mechanism they are intended to study whenever possible. Deterministic generation and matched seeds cannot isolate a mechanism if a variant changes several important parts of the workflow at once.

Third, benchmark designers should verify that agents actually reach the states needed for the intended comparison. A task branch provides little evidence about a behavior if evaluated agents consistently leave the path before reaching that branch.

Finally, interventions should be treated as diagnostic probes. The cross-generation reversal shows that guidance that helps one model may affect another differently. Its purpose is to test a specific explanation.

\subsection{Reproducibility and Analysis Discipline}

The cross-generation reversal also shows why experimental controls matter. If prompts, interventions, execution order, or analysis rules can be changed after results are observed, it becomes difficult to know whether an unexpected finding reflects model behavior or changes in the experiment.

In our confirmatory studies, the experimental design and analysis decisions are fixed before the corresponding outcomes are examined, and infrastructure failures remain part of the experiment record rather than being silently replaced. This was especially important for the Gemini 3.7 replication: the intervention was kept unchanged even though it produced the opposite effect from the Gemini 2.5 study. Preserving unexpected and negative results is necessary for detecting genuine changes in model behavior.

Our conclusions are limited to the controlled settings studied here. The tasks are designed to isolate specific long-horizon failure mechanisms rather than measure general autonomous penetration-testing ability, and the intervention results apply to the evaluated model--task combinations. The Gemini 3.7 study shows that the earlier discovery diagnosis no longer transfers in the same way, but it does not determine the specific cause of the later downstream failures.

\section{Conclusion}
\label{sec:conclusion}

This paper presents a diagnostic methodology for understanding why long-horizon security agents fail. Instead of relying only on end-to-end success, we use checkpoints to determine whether an agent reaches the point where a target capability can actually be exercised. We then examine what happens after that point and use controlled rescue--placebo interventions to test suspected upstream failure mechanisms.

Across four diagnostic task families, this approach shows that final success can hide where failures occur. In Controlled State Reuse, many Gemini 2.5 Flash failures happen before the model reaches the state needed for later reuse, and a controlled intervention shows that protocol discovery contributes to this bottleneck. Repeating the same frozen intervention with Gemini 3.7 Flash produces the opposite effect, while reaching the same state no longer reliably leads to task completion. Thus, the main source of failure can change across model generations even when the task and experimental design remain fixed.

These results motivate evaluating long-horizon agents not only by whether they finish a task, but also by whether they reach the states needed to exercise the capabilities being tested and where failures occur afterward. Controlled interventions can then test whether a suspected obstacle actually contributes to those failures. These tools provide a clearer way to distinguish failures of discovery and exposure from failures of the downstream capability itself.

\section{Ethical Considerations}
\label{sec:ethics}

This work evaluates offensive-security agents in controlled, deterministic Docker environments rather than against production or third-party systems. Task secrets, configuration values, and other security-relevant state are generated within the benchmark, and the interventions are designed to diagnose agent failure mechanisms rather than directly solve the tasks.

The work nevertheless has dual-use potential because better understanding agent failures could also improve offensive agents. We therefore focus on evaluation, diagnosis, and benchmark design, and report unexpected results---including the Gemini 3.7 intervention reversal---rather than modifying experiments to obtain a desired outcome.

\bibliographystyle{IEEEtran}
\bibliography{ref.bib}

\end{document}